\documentclass[conference,compsoc]{IEEEtran}
\IEEEoverridecommandlockouts

\usepackage{cite}
\usepackage{amsmath,amssymb,amsfonts}
\usepackage{algorithmic}
\usepackage{graphicx}
\usepackage{textcomp}
\usepackage{xcolor}
\usepackage[utf8]{inputenc}
\usepackage[T1]{fontenc}
\usepackage{hyperref}
\usepackage{url}
\usepackage{cleveref}
\usepackage{booktabs}
\usepackage{amsfonts}
\usepackage{nicefrac}
\usepackage{microtype}
\usepackage{lipsum}
\usepackage{graphicx}
\usepackage{xcolor}
\usepackage{booktabs}
\graphicspath{ {./images/} }

\newcommand{\subheading}[1]{\noindent{\textbf{#1}}}

\def\BibTeX{{\rm B\kern-.05em{\sc i\kern-.025em b}\kern-.08em
    T\kern-.1667em\lower.7ex\hbox{E}\kern-.125emX}}
\begin{document}


\title{Can Current Detectors Catch Face-to-Voice Deepfake Attacks?}

\author{\IEEEauthorblockN{Nguyen Linh Bao Nguyen}
\IEEEauthorblockA{
\textit{CSIRO's Data61}\\
Melbourne, Australia \\
bao.nguyen@csiro.au}
\and
\IEEEauthorblockN{Alsharif Abuadbba}
\IEEEauthorblockA{
\textit{CSIRO's Data61}\\
Sydney, Australia \\
sharif.abuadbba@csiro.au}
\and
\IEEEauthorblockN{Kristen Moore}
\IEEEauthorblockA{ 
\textit{CSIRO's Data61}\\
Melbourne, Australia \\
kristen.moore@csiro.au}
\and
\IEEEauthorblockN{Tingmin Wu}
\IEEEauthorblockA{ 
\textit{CSIRO's Data61}\\
Melbourne, Australia \\
tina.wu@@csiro.au}

}

\maketitle

\begin{abstract}

The rapid advancement of generative models has enabled the creation of increasingly stealthy synthetic voices, commonly referred to as audio deepfakes. A recent technique, FOICE [USENIX’24], demonstrates a particularly alarming capability: generating a victim’s voice from a single facial image, without requiring any voice sample. By exploiting correlations between facial and vocal features, FOICE produces synthetic voices realistic enough to bypass industry-standard authentication systems, including WeChat Voiceprint and Microsoft Azure. This raises serious security concerns, as facial images are far easier for adversaries to obtain than voice samples, dramatically lowering the barrier to large-scale attacks.
In this work, we investigate two core research questions: (\textbf{RQ1}) can state-of-the-art audio deepfake detectors reliably detect FOICE-generated speech under clean and noisy conditions, and (\textbf{RQ2}) whether fine-tuning these detectors on FOICE data improves detection without overfitting, thereby preserving robustness to unseen voice generators such as SpeechT5. 

Our study makes three contributions. First, we present the first systematic evaluation of FOICE detection, showing that leading detectors consistently fail under both standard and noisy conditions. Second, we introduce targeted fine-tuning strategies that capture FOICE-specific artifacts, yielding significant accuracy improvements. Third, we assess generalization after fine-tuning, revealing trade-offs between specialization to FOICE and robustness to unseen synthesis pipelines. These findings expose fundamental weaknesses in today’s defenses and motivate new architectures and training protocols for next-generation audio deepfake detection.
\end{abstract}


\section{Introduction}
\label{intro}

Recent advances in generative AI have enabled speech synthesis models capable of mimicking human voices with remarkable realism. While these systems have legitimate uses in entertainment and accessibility, they also introduce serious risks to security, privacy, and trust. Synthetic voices are increasingly used to impersonate individuals, bypass authentication systems, and facilitate fraud\cite{survey}. To counter these risks, researchers have developed a growing body of audio deepfake detectors~\cite{blue2022usenix,kumari2025voiceradar, sun2023aisynth}. However, existing research on audio deepfakes has largely focused on text- or voice-conditioned pipelines, where artifacts left by vocoders provide detectable cues~\cite{yadav2024fairssd, yan2025voicewukong}. The emergence of new cross-modal synthesis methods has begun to erode these assumptions.

\begin{figure}[t]
    \centering
    \includegraphics[width=0.5\textwidth]{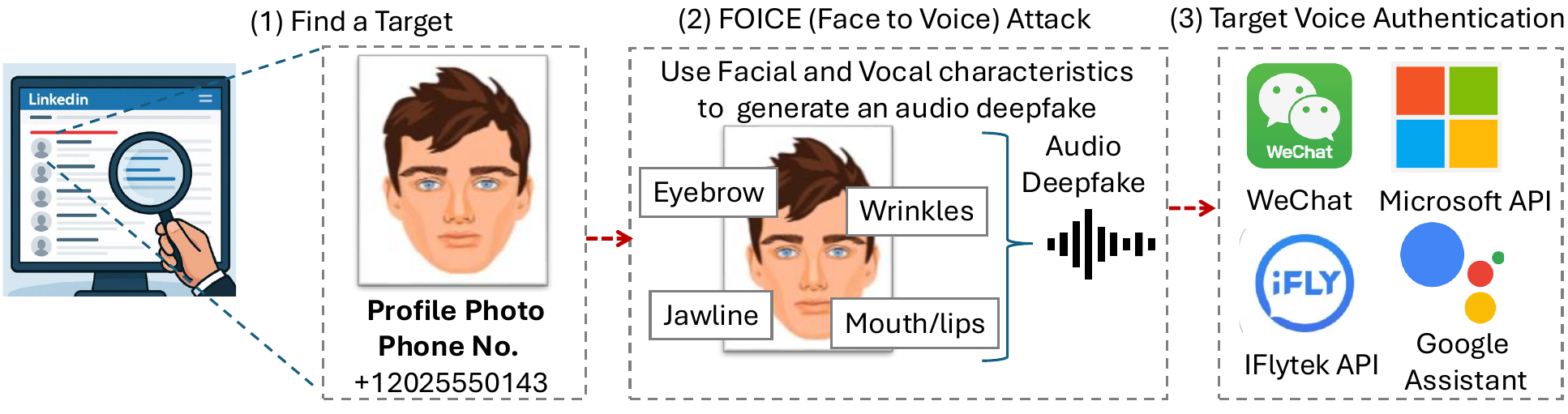}
    \caption{Illustration of Face to Voice Deepfake}
    \label{fig:foice_attack}
\end{figure}

A particularly concerning example is \textbf{Face-to-Voice (FOICE) synthesis} \cite{FOICE}, which generates a synthetic voice from a single facial image by exploiting correlations between facial and vocal characteristics, as illustrated in Figure~\ref{fig:foice_attack}. 
This enables large-scale impersonation attacks without needing any clean voice samples, since facial images are readily available on social media. For example, \textit{an attacker could scrape a victim’s photo and phone number (often used as a WeChat ID) from social media such as LinkedIn and, using an SV2TTS‑style vocoder, render attacker‑supplied text (e.g., authentication digits) in a voice closely resembling the target, producing speech convincing enough to bypass authentication systems like WeChat Voiceprint and Microsoft Azure}. These capabilities fundamentally shift the threat landscape: attackers are no longer constrained by voice data availability, while existing detectors remain optimized for vocoder-driven synthesis.

Despite the severity of this risk, no prior work has systematically evaluated how current audio deepfake detectors perform against cross-modal attacks such as FOICE. In this paper, we fill that gap through a comprehensive evaluation of state-of-the-art detectors under clean and noisy conditions, followed by targeted fine-tuning to test adaptation and generalization. We focus on two key questions:

\begin{itemize}
    \item \textbf{RQ1 (Baseline Detection on FOICE).} Can current state-of-the-art detectors reliably detect FOICE-generated speech, including under realistic noise conditions?
    \item \textbf{RQ2 (Fine-tuning vs. Overfitting).} Can fine-tuning detectors on FOICE improve performance without overfitting - i.e., while preserving robustness to unseen pipelines such as SpeechT5?
\end{itemize}

To answer these questions, we make the following contributions:

\begin{itemize}
    \item \textbf{Systematic evaluation (RQ1).} We conduct the first empirical study of FOICE detection, revealing that all tested detectors fail to reliably identify FOICE-generated audio across clean, noisy, and de-noised conditions.
    \item \textbf{Fine-tuning for mitigation (RQ2).} We introduce targeted fine-tuning protocols that significantly improve FOICE detection across multiple architectures.
    \item \textbf{Generalization analysis (RQ2).} We evaluate fine-tuned detectors against the unseen SpeechT5 pipeline, revealing trade-offs between specialization and cross-domain robustness.
\end{itemize}

Together, these findings highlight the urgent need to rethink audio deepfake defenses beyond vocoder-centric assumptions, emphasizing architectures and training protocols that can adapt to emerging cross-modal synthesis methods such as FOICE.

\section{Background and Related Work}

\subsection{Audio Deepfakes and Detection Landscape}

The problem of detecting synthetic speech has been studied extensively in recent years. Early work focused on handcrafted spectral, phase, and prosodic features combined with machine learning classifiers, formalized through shared tasks such as the ASVspoof challenges \cite{wang2020asvspoof}. These benchmarks remain central for evaluation, but detectors trained solely on them often inherit dataset-specific biases. Subsequent corpora, such as WaveFake \cite{wavefake}, LibriSeVoc \cite{AISVGeneralization}, and FakeAVCeleb \cite{fakeavceleb}, broadened vocoder and synthesis diversity, yet cross-domain generalization remains limited.  

Modern detectors increasingly adopt deep neural architectures. Convolutional and recurrent models applied to spectrograms, as well as waveform-level networks, dominate the landscape. More recent designs leverage attention or graph-based modeling to capture both local and global structure, with AASIST \cite{AASIST} as a representative state-of-the-art detector. Parallel efforts explicitly target artifacts introduced during speech synthesis, e.g., residual noise or vocoder frequency patterns \cite{VocoderArtifacts}, while others explore domain-invariant representations \cite{AISVGeneralization} and pretraining with large-scale multimodal encoders \cite{llm_deepfake}. Temporal–channel modeling \cite{tcm_add} has also been investigated to improve robustness against noisy playback and unseen vocoder types.  

Despite these advances, most detectors remain tuned to the artifact patterns of vocoder-based text-to-speech or voice-conversion pipelines, limiting their ability to generalize to alternative generation mechanisms. As a result, detectors degrade sharply when confronted with unseen generation methods, such as SpeechT5 \cite{speecht5}, or under realistic environmental noise conditions. Crucially, \textit{no existing benchmark or study has systematically evaluated detection against face-conditioned, cross-modal synthesis such as FOICE.}

\subsection{FOICE and Face-to-Voice Synthesis}

FOICE, introduced at USENIX Security 2024 \cite{FOICE}, exemplifies a qualitatively new and stealthier threat model. Instead of relying on text or voice conditioning, FOICE infers how a person would sound directly from a \emph{single facial image}. As shown in Figure~\ref{fig:foice_attack}, it exploits the biological correlation between facial structure (e.g., jawline, lips, bone morphology) and vocal attributes (e.g., pitch, timbre, resonance). The model separates \emph{face-dependent} voice features from \emph{face-independent} features sampled from a latent Gaussian space. These are fused and synthesized through a SV2TTS-style vocoder, which renders attacker-provided text in the inferred voice. This design makes FOICE stealthy and scalable: facial images are far easier to obtain than voice recordings, and FOICE has been shown to succeed even from low-resolution or partially occluded faces~\cite{FOICE}. Because FOICE is conditioned on facial rather than vocal inputs, its outputs lack many of the frequency-domain artifacts that detectors rely on, creating a serious blind spot for current defenses. In evaluations, FOICE achieved average per-victim bypass success rates of $\sim$30\% against WeChat Voiceprint and nearly 100\% overall success when multiple attempts were allowed \cite{FOICE}. 

\subsection{Threat Model and Research Gap}

The FOICE attack introduces a novel adversarial setting that departs significantly from assumptions underlying most prior defences.  
\textbf{Attacker capabilities:} The adversary requires only a single facial image of the target (readily obtainable from public sources sucha as social media) and a trained face-to-voice synthesis model. Auxiliary metadata such as a phone number (e.g., a WeChat ID) may also be leveraged to target specific platforms.   
\textbf{Victim assets:} The primary targets include biometric authentication systems (e.g., WeChat Voiceprint, Microsoft Azure), Know-Your-Customer (KYC) APIs used in financial platforms, and the broader trustworthiness of voice as a biometric identity signal.  
\textbf{Attack goals:} The adversary seeks to gain unauthorized access to protected accounts, commit fraud at scale, impersonation users in sensitive government or financial contexts, or inject falsified speech into legal, journalistic, or evidentiary workflows.  

From a defense perspective, FOICE presents a critical blind spot. Its outputs are not conditioned on speech content or victim voice samples, unlike text-to-speech or voice conversion pipelines, meaning detectors trained on vocoder-specific artifacts may fail to generalize. While prior research has explored robustness to unseen vocoders and playback distortion, no existing system has been systematically evaluated against cross-modal synthesis methods such as FOICE. This gap motivates our study, which investigates two key research questions: (\textbf{RQ1}) whether current detectors can identify FOICE-generated speech, and (\textbf{RQ2}) whether fine-tuning can improve detection without compromising generalization to other pipelines.

\section{Methodology}
In this section, we present our systematic evaluation framework for assessing whether state-of-the-art audio deepfake detectors can reliably detect FOICE-generated speech and whether fine-tuning improves robustness without sacrificing generalization.

\subsection{Overview}

Our evaluation proceeds in three main stages:  
(1) \textbf{Models:} We benchmark diverse detection architectures using their unmodified, author-released checkpoints to ensure fairness and reproducibility.  
(2) \textbf{Datasets:} We construct a balanced corpus of real and synthetic speech samples, including FOICE-generated and out-of-distribution fakes, under clean, noisy, and denoised conditions.  
(3) \textbf{Fine-tuning:} We implement a structured fine-tuning mechanism to integrate FOICE-specific data and tests for adaptation versus catastrophic forgetting on unseen pipelines.

\subsection{Detection Models}

We evaluate four representative detection models that span distinct architectural families and design goals. All models are initialized from publicly released checkpoints.

\subheading{(1) AASIST} \cite{AASIST}. A waveform-level model that combines spectro–temporal front-ends with graph attention to capture both local and global dependencies.  Its architecture makes it a strong candidate for detecting cross-modal synthesis artifacts beyond traditional vocoder traces.

\subheading{(2) Ren et al.} \cite{AISVGeneralization}. A model designed to separate domain-specific from domain-invariant features to support cross-domain generalization. We expect this model to be informative for testing resilience to unseen pipelines (e.g., SpeechT5) and to resist catastrophic forgetting under FOICE fine-tuning.

\subheading{(3) Sun et al.} \cite{VocoderArtifacts}. A spectrogram-based model trained to identify residual cues left by neural vocoders. Although effective on vocoder-based synthesis, it serves here as a stress-test for FOICE, which lacks many such artifacts.

\subheading{(4) Temporal–Channel Modeling (TCM)} \cite{tcm_add}. A self-attention-based model that captures by attending jointly over temporal and channel dimensions. It provides a strong attention-based baseline under both in-distribution and out-of-distribution conditions.

\subsection{Datasets}

To evaluate FOICE detection performance, we construct a dataset combining real, cross-modal, and vocoder-driven speech from four sources: 

\subheading{(A) VoxCeleb2 \cite{chung2018voxceleb2}.} Contains over one million utterances from thousands of celebrity speakers recorded in diverse recording environments. Used both to generate FOICE voices and to evaluate robustness on in-distribution real speech. 

\subheading{(B) AVSpeech \cite{ephrat2018looking}.} Comprises large-scale audio-visual speech segments from YouTube, with cleaner acoustic conditions and more diverse (non-celebrity) speaker identities. Enables evaluation of generalization beyond FOICE’s celebrity-centric training distribution.  

\subheading{(C) FOICE.} We generate synthetic samples by conditioning FOICE on facial images from VoxCeleb2 and AVSpeech, producing paired real–fake audio  with matched and mismatched speaker distributions.  

\subheading{(D) SpeechT5 DS \cite{speecht5}.}  Includes 1,400 synthetic samples generated using SpeechT5 (which leverages HiFi-GAN~\cite{hifi}). Used exclusively at evaluation time to test cross-domain generalization (RQ2); it is excluded from baseline training and FOICE fine-tuning stages.

Overall, the dataset comprises 15,144 samples: 6,261 labeled as fake and 8,883 as real. All real speech comes from VoxCeleb2 and AVSpeech, while synthetic samples are split between FOICE and SpeechT5. The dataset is balanced across three playback conditions: clean, noisy (via \texttt{audiomentations}~\cite{audiomentations}), and denoised (via \texttt{noisereduce}~\cite{noisereduce}). Real and fake distributions are matched across noise settings to ensure robust evaluation.

\subsection{Fine-Tuning Mechanism}\label{sec:fine_tuning}

We introduce a controlled fine-tuning protocol to explore the trade-off between adaptation to FOICE and retention of cross-domain generalization.  Below are the details for the training design, optimisation strategies we followed and goals. 

\subheading{Training Design.} Mini-batches interleave lean, noisy, and denoised samples to ensure robust gradient signals. Regularization is enforced through weight decay and cosine learning-rate schedules. Models are selected based on held-out validation accuracy rather than early stopping, ensuring consistent comparisons.  

\subheading{Optimization.} We retain the hyperparameter settings recommended by the original authors, making minimal modifications for dataset scale. For example:  

\subheading{(1) AASIST \cite{AASIST}:} batch size 24, 10 epochs, Adam with $\beta=(0.9,0.999)$, CCE loss, base LR $1\times10^{-4}$, weight decay $1\times10^{-4}$.  

\subheading{(2) Ren et al. \cite{AISVGeneralization}:} batch size 8, 30 epochs, cosine annealing, initial LR per original paper, weight decay $1\times10^{-4}$.  

\subheading{(3) Sun et al. \cite{VocoderArtifacts}:} batch size 32, 5 epochs, LR $1\times10^{-4}$, weight decay $1\times10^{-4}$. 

\subheading{(4) TCM \cite{tcm_add}:} batch size 16, 30 epochs, LR $1\times10^{-6}$, weight decay $1\times10^{-4}$. 

\subheading{Goals.} Fine-tuning serves two complementary purposes:
(1) \textit{Adaptation:} Improves baseline detection by incorporating FOICE-specific cues.  
(2) \textit{Resilience:} Assess whether fine-tuned detectors retain performance on unseen pipelines such as SpeechT5, or whether they suffer catastrophic forgetting.  

This framework enables a direct evaluation of the trade-off between specialization to FOICE and generalization to future, unseen deepfake synthesis methods.

\subsection{Experimental Setup}
Our evaluation pipeline comprises preprocessing, training, and evaluation phases. We conduct all experiments on NVIDIA H100 GPUs with 200GB RAM.

\textbf{Preprocessing:} Each audio sample is prepared in three forms—clean, noisy, and denoised—to reflect real-world playback conditions. Noise is injected using Gaussian perturbations via \texttt{audiomentations}, while denoising is applied using spectral gating via \texttt{noisereduce}.

\textbf{Training:} We construct balanced 80/20 train–test splits across real and fake classes, ensuring speakers are strictly disjoint to prevent identity leakage. Fine-tuning uses the procedure in Section~\ref{sec:fine_tuning}, with hyperparameters aligned to original implementations. The fine-tuning set contains around 4,200 FOICE fakes while real examples are more numerous, producing a class imbalance. Because the majority is real, adaptation risks under-sensitivity to novel fakes rather than a “predict-all-fake” failure. We therefore report threshold-agnostic (EER) and class-aware metrics (precision, recall, F1), include confusion matrices, and select checkpoints on a balanced validation split to mitigate threshold distortion.

\textbf{Evaluation:} Models are tested on both in-distribution (FOICE) and out-of-distribution (SpeechT5 with HiFi-GAN) audio. We report standard classification metrics, including accuracy, precision, recall, F1-score, specificity, and Equal Error Rate (EER). EER represents the operating point where false acceptance and rejection rates are equal; lower EER indicates better performance. Recall is especially important in our setting, as missed detections may result in security breaches.

\section{Evaluation}

We now present our experimental evaluation of state-of-the-art audio deepfake detectors against FOICE attacks. Our study aims to (i) assess baseline detector performance on FOICE data (RQ1), and (ii) determine whether fine-tuning improves detection while maintaining robustness on unseen pipelines such as SpeechT5 (RQ2).

\subheading{RQ1 — Baseline Detection on FOICE.}  
Baseline detectors (denoted "Base" in the Model column) show limited discriminative ability on FOICE-generated speech, with high EERs and inconsistent accuracy across clean, noisy, and denoised audio, as shown in Table~\ref{tab:combined_results}, Figures~\ref{fig:eer_combined_comparison} and \ref{fig:accuracy_combined_comparison} in Appendix~\ref{appendix:accuracy}.

The vocoder artifact detector by Sun et al.~\cite{VocoderArtifacts} and TCM are the weakest performers: Sun et al.~\cite{VocoderArtifacts} underperforms due to reliance on vocoder residue cues that FOICE does not expose reliably, while TCM exhibits systematic over-detection, resulting in inflated recall at the expense of accuracy, precision, F1-Score across clean, noisy, and denoised conditions. Ren et al.’s model~\cite{AISVGeneralization} offers the most stable cross-condition performance and achieves the highest F1-score, suggesting strong alignment with the signal characteristics of FOICE speech. AASIST, while slightly behind in F1, maintains a solid balance of sensitivity and specificity, outperforming artifact-centric models on robustness across conditions.

\begin{figure*}[ht]
    \centering
    \includegraphics[width=0.75\textwidth]{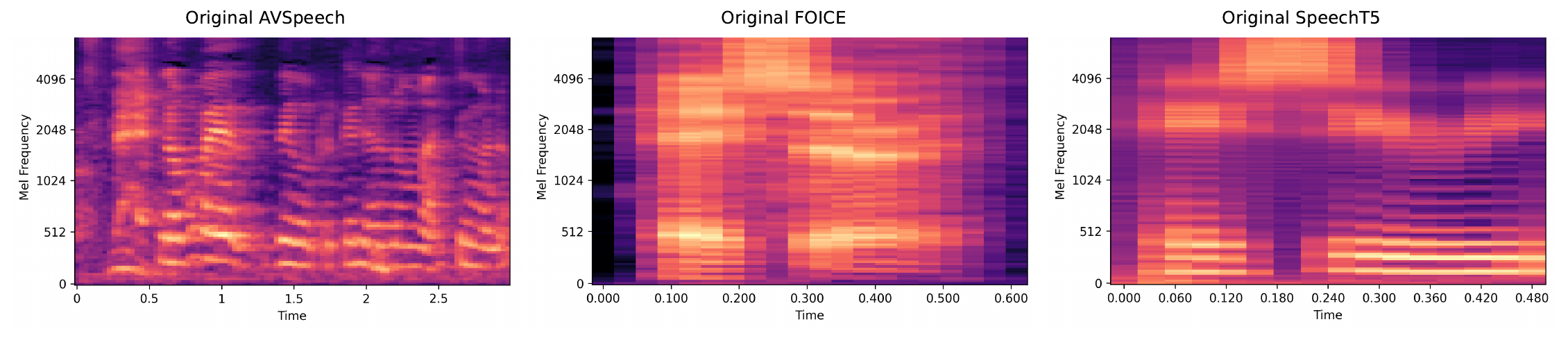}
    \caption{Mel spectrograms comparing audio from three pipelines: (left) AVSpeech (real), (middle) FOICE generated with the SV2TTS vocoder, and (right) SpeechT5 generated with HiFi-GAN vocoder. Each spectrogram shows distinct time–frequency signatures and synthesis artifacts. FOICE, in particular, exhibits banding and smoothing effects absent from AVSpeech’s richer harmonic structures, illustrating how cross-vocoder differences create challenges for robust deepfake detection.}
    \label{fig:spectrogram}
\end{figure*}

These results demonstrate that FOICE introduces cross-modal artifacts absent from standard benchmarks such as ASVspoof or WaveFake. Because FOICE is not conditioned on transcripts or clean voice samples, detectors trained on vocoder-driven pipelines lack the discriminative features needed to flag such attacks, exposing a blind spot in current defenses. 
This phenomenon is visually supported by the spectrogram comparisons in Figure~\ref{fig:spectrogram}, which compares a real speech sample from AVSpeech, a FOICE-generated sample synthesized using the SV2TTS vocoder, and a SpeechT5-generated sample synthesized with HiFi-GAN. These comparisons emphasize unique frequency bands and synthesis artifacts characteristic of each source. Further spectrograms in Appendix~\ref{appendix:other_spectrograms} highlight vocoder-specific patterns in spectrograms of each detector's original training data.

\begin{table*}[h]
\centering
\caption{Detector Performance Metrics. For each metric, arrows indicate whether a higher score is preferred (↑) or not (↓): EER (↓), Accuracy (↑), Precision (↑), Recall (↑), F1-Score (↑).}
\label{tab:combined_results}
\begin{tabular}{llcccccccccc}
\toprule
\textbf{Detector} & \textbf{Model} & \textbf{EER (↓)} & \textbf{Accuracy (↑)} & \textbf{Precision (↑)} & \textbf{Recall (↑)} & \textbf{F1-Score (↑)} & \textbf{TP} & \textbf{TN} & \textbf{FP} & \textbf{FN} \\
\midrule
AASIST~\cite{AASIST} & Base & 0.163 & 0.735 & 0.416 & 1.000 & 0.587 & 687 & 1996 & 965 & 0 \\
AASIST~\cite{AASIST} & Fine-tuned & 0.003 & 0.995 & 0.972 & 1.000 & 0.986 & 687 & 2941 & 20 & 0 \\
Ren et al.~\cite{AISVGeneralization} & Base & 0.271 & 0.898 & 1.000 & 0.457 & 0.627 & 314 & 2961 & 0 & 373 \\
Ren et al.~\cite{AISVGeneralization} & Fine-tuned & 0.002 & 0.998 & 0.987 & 1.000 & 0.993 & 687 & 2952 & 9 & 0 \\
Sun et al.~\cite{VocoderArtifacts} & Base & 0.566 & 0.630 & 0.099 & 0.119 & 0.108 & 82 & 2213 & 745 & 605 \\
Sun et al.~\cite{VocoderArtifacts} & Fine-tuned & 0.004 & 0.995 & 0.974 & 0.999 & 0.986 & 686 & 2940 & 18 & 1 \\
TCM~\cite{tcm_add} & Base & 0.433 & 0.297 & 0.211 & 1.000 & 0.349 & 687 & 395 & 2563 & 0 \\
TCM~\cite{tcm_add} & Fine-tuned & 0.150 & 0.913 & 0.779 & 0.750 & 0.764 & 515 & 2812 & 146 & 172 \\
\bottomrule
\end{tabular}
\end{table*}

\begin{figure*}[t]
    \centering
    \includegraphics[width=0.85\textwidth]{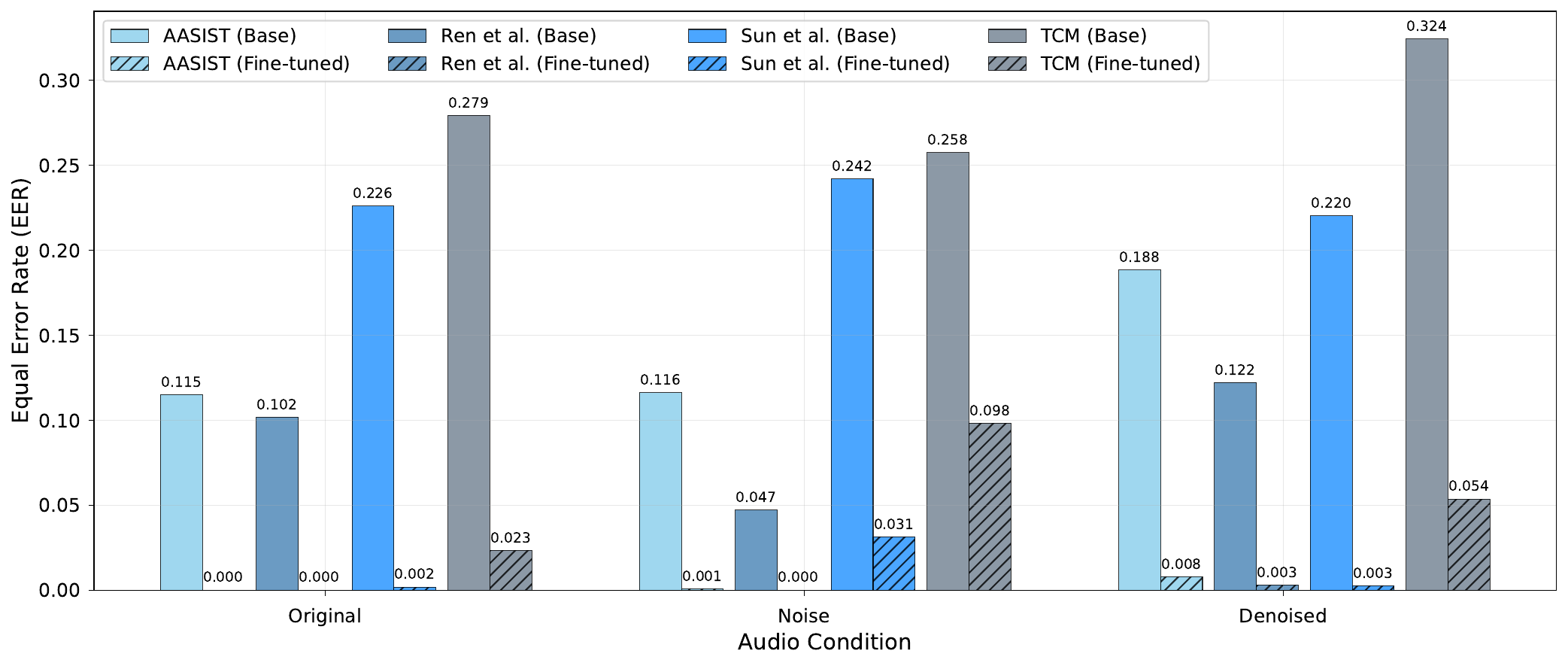}
    \caption{Equal Error Rate (EER) of the four detectors—AASIST, Ren et al., Sun et al., and TCM—on FOICE data under three audio conditions (original, noisy, denoised). Bars show baseline (solid) and FOICE-fine-tuned (hatched) models. Lower EER indicates stronger discrimination between real and fake audio. Fine-tuning yields large reductions in EER across all conditions, with AASIST and Ren et al. approaching near-zero values. TCM shows the least improvement but still benefits.}
    \label{fig:eer_combined_comparison}
\end{figure*}

\begin{figure}[ht]
    \centering
    \includegraphics[width=0.5\textwidth]{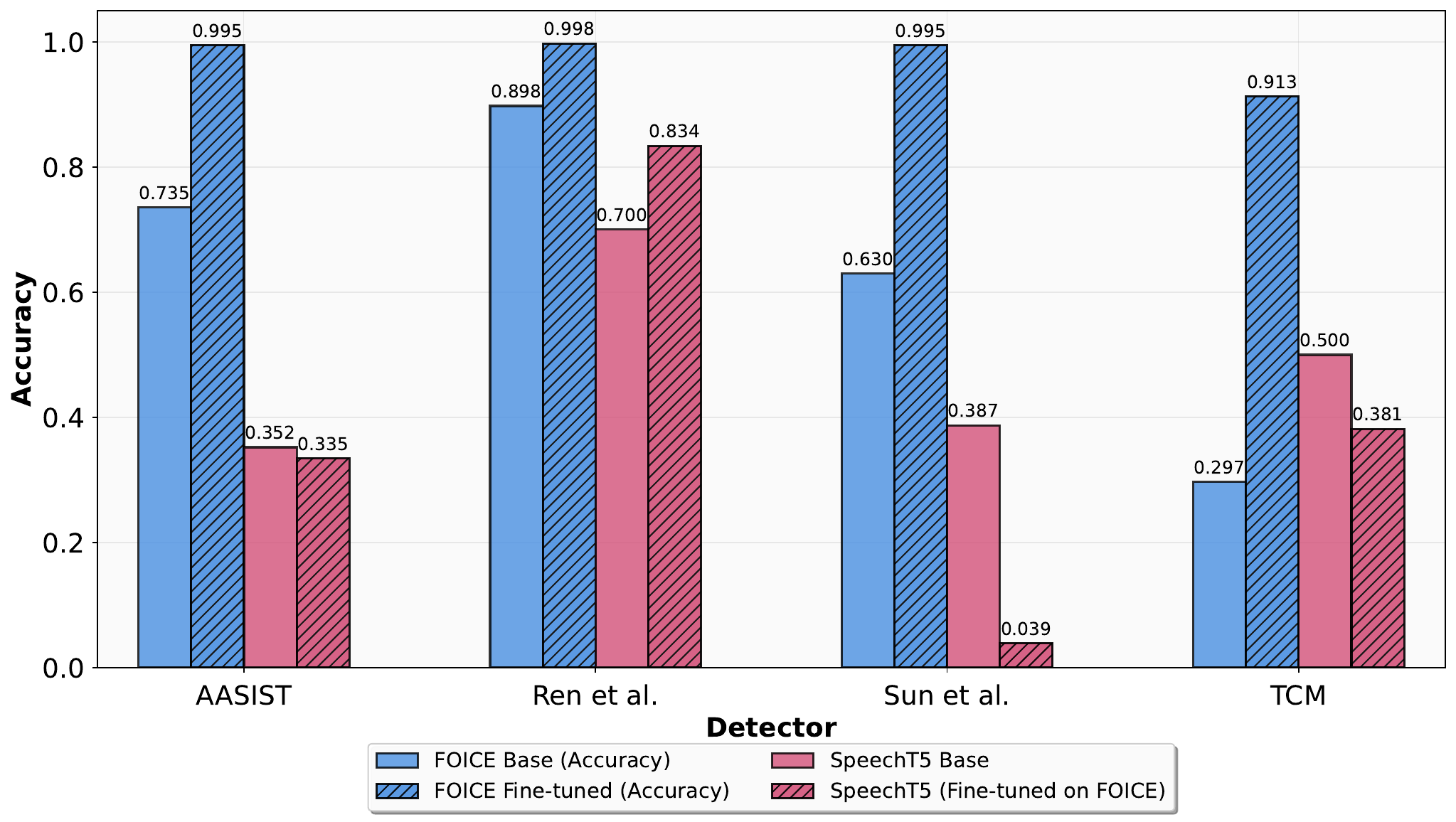}
    \caption{Accuracy of four detectors under baseline (solid) and FOICE‑fine‑tuned (hatched) regimes across two pipelines: FOICE with SV2TTS (blue) and SpeechT5 with HiFi‑GAN (red). The results highlight a trade‑off between in‑distribution adaptation and generalization to unseen synthesis methods when optimizing for accuracy.}
    \label{fig:foice_vs_speecht5_accuracy_comparison}
\end{figure}

\subheading{RQ2 — Fine-tuning vs. Overfitting.}  
Fine-tuning on FOICE data yields substantial performance gains across all detectors. Previously underperforming models such as TCM and Sun et al. \cite{VocoderArtifacts} achieve near-perfect accuracy after fine-tuning, as seen in Table~\ref{tab:combined_results} (Model = "Fine-tuned"). Even the stronger models AASIST and Ren et al.~\cite{AISVGeneralization} improve, suggesting that prior failures stemmed from gaps in training data rather than fundamental architectural limitations.  
However, these gains often come at the expense of generalization. When evaluated on the out-of-distribution SpeechT5 dataset, fine-tuned models experience significant accuracy drops, as shown in Figure~~\ref{fig:foice_vs_speecht5_accuracy_comparison}. AASIST maintains a stable accuracy while TCM drops by nearly 10\%. Sun et al.~\cite{VocoderArtifacts}'s detector nearly collapses (dropping from 38.7\% to 3.9\% accuracy). In contrast, only Ren et al.~\cite{AISVGeneralization}'s domain invariant model improves, demonstrating the advantage of architectures explicitly designed for balancing adaptation and generalization.  

This trade-off highlights a persistent vulnerability in audio forensics: while detectors can rapidly adapt to FOICE-specific cues, such specialization can lead to overfitting and sometimes catastrophic forgetting, leaving systems brittle to even modest changes in synthesis pipelines.

\subheading{Key Takeaways.} Our findings underscore two critical points. First, state-of-the-art detectors exhibit clear blind spots when confronted with FOICE deepfakes, which fall outside the assumptions of vocoder-based synthesis pipelines. Second, while fine-tuning can yield near-perfect detection within-distribution, it often erodes robustness against unseen synthesis, unless the model architecture is explicitly optimized for domain robustness, as demonstrated by Ren et al.~\cite{AISVGeneralization}. Addressing this trade-off is essential for building durable defenses against evolving cross-modal deepfake techniques.

\section{Discussion}

Our study shows that FOICE-generated speech exposes a significant blind spot in current detection systems: detectors trained on traditional vocoder pipelines fail out-of-the-box, can be rapidly adapted via fine-tuning, but often lose robustness to unseen synthesis methods.\\

\subheading{Security implications.} FOICE signals a fundamental shift in audio deepfake generation, from text- or voice-conditioned synthesis to face-conditioned pipelines. Unlike prior methods that rely on clean audio or transcripts, FOICE requires only a single facial image, readily obtainable from public social media, to produce convincing speech. Moreover, the resulting audio lacks common vocoder artifacts that existing detectors rely on, allowing attackers to bypass models optimized for traditional synthesis traces. As face-to-audio generation matures, its potential to undermine biometric authentication systems—especially those relying on voice verification—presents an urgent challenge beyond traditional media forensics.

\subheading{Fine-tuning as a first response.} Encouragingly, we find that state-of-the-art detectors can be quickly adapted to FOICE via fine-tuning, achieving high accuracy on in-distribution samples with modest data. This suggests that, when new synthesis methods emerge, organizations can deploy reactive defenses without overhauling their infrastructure. However, these benefits are fragile: fine-tuned detectors can overfit to FOICE-specific artifacts, showing sharp performance degradation on unseen pipelines like SpeechT5. This reveals an inherent trade-off between specialization and robustness—a key limitation of current methods.

This fine-tuning–generalization trade-off mirrors long-standing challenges observed in the visual deepfake detection literature. Numerous studies show that detectors trained on specific generators or datasets rarely generalize to unseen manipulations, often overfitting to superficial artifacts or identity-specific cues. For example, Le et al. (EuroS\&P 2025)~\cite{le2025sok} show that none of sixteen state-of-the-art detectors maintained robust cross-generator performance, several even performed worse than random guessing on novel forgeries. Yang et al. (CVPR 2025)~\cite{yang2025d} similarly demonstrate that detectors optimized on a single generator exhibit poor transfer to others, and propose multi-generator discrepancy learning to improve robustness. Dong et al. (CVPR 2023)~\cite{Dong_2023_CVPR} further attribute this fragility to implicit identity leakage, where detectors inadvertently entangle artifact detection with identity cues, limiting cross-domain effectiveness. Collectively, these findings reinforce our results on FOICE: while fine-tuning offers strong in-distribution performance, its specialization often comes at the cost of generalization, a systemic limitation shared across both audio and visual modalities.

\subheading{Error patterns across detectors.}
An inspection of confusion matrix counts reveals divergent failure modes: baseline TCM misclassifies nearly all real samples as fake, leading to extremely high false positive rates, while base Ren et al.~\cite{AISVGeneralization} exhibits the opposite pattern with widespread false negatives. These biases underscore that detectors not only vary in performance but also in the nature of their errors, highlighting the importance of examining decision tendencies beyond aggregate metrics.

\subheading{Detection under noisy and denoised conditions.}
Our experiments also highlight the impact of noise and audio preprocessing on detection performance. Interestingly, we observe that denoised audio is often more challenging for detectors than noisy audio. While fine-tuned AASIST and Sun et al.~\cite{VocoderArtifacts} models remain relatively stable across both conditions, detectors like TCM and Ren et al.\cite{AISVGeneralization} show substantial degradation post-denoising. TCM’s fine-tuned variant, for example, sees accuracy drop below 90\% and EER spike to 0.324 on denoised audio. This may suggest that certain detectors rely on fine-grained noise artifacts or phase distortions that are inadvertently removed by denoising algorithms, masking cues needed for detection. Future work should explore whether more adaptive or perceptually-aligned denoising techniques can retain deepfake-relevant artifacts without suppressing critical detection signals.

\subheading{Deployment challenges and next steps.} While our evaluation covers key scenarios (clean, noisy, denoised) it remains intentionally scoped and therefore limited. We did not evaluate FOICE combined with alternative vocoders (e.g., DiffWave, MelGAN, WaveGrad) or with video-derived conditioning, which may introduce new artifacts or compound generalization challenges. Real-world deployments also involve variable playback channels, compression, device heterogeneity, and adversarial preprocessing that are not fully captured here. Finally, relying on fine-tuning as a first response is necessarily reactive; proactive, deployment-ready defenses remain an open problem.

To address these gaps, we propose a focused research agenda. First, researchers should expand generator coverage by evaluating FOICE variants integrated with alternative vocoders and video-conditioned generators to measure how vocoder choice and multi-modal conditioning affect both attack success and detector transfer. Second, studies must examine robustness under realistic channels, testing detectors across a matrix of codecs, microphone types, playback devices, and network transport conditions (including compression and packet loss) to better approximate operational environments. Finally, we encourage focused mitigation research on multi-generator / discrepancy training (cf. ~\cite{yang2025d}), domain-invariant representation learning, class-imbalance–aware fine-tuning, and perceptually-aware denoising that preserves forensic cues. These directions directly address the limitations above and point toward defensible, deployment-oriented research priorities.


\section{Conclusion}

We present the first systematic evaluation of audio deepfake detectors against FOICE, a face-conditioned synthesis method that departs from traditional vocoder-driven pipelines. Evaluating four state-of-the-art detectors across (i) in-distribution FOICE detection, (ii) generalization to unseen generators (SpeechT5), and (iii) noisy and denoised conditions, we find that fine-tuning yields strong in-distribution performance but often erodes robustness to novel synthesis methods. Only domain-invariant approaches maintained relatively stable cross-vocoder behavior; noise robustness varied widely, and denoising can unintentionally remove forensic cues. Lasting defenses therefore require (i) larger, more diverse corpora (including FOICE variants and modern vocoders) and (ii) architectures and training regimes that target vocoder-independent, cross-modal representations. By identifying where current detectors succeed and fail, our study aims to guide development of more robust, generalizable deepfake audio defenses.

\bibliography{references.bib}
\bibliographystyle{plain}

\appendices
\section{Detectors Accuracy on FOICE Data}\label{appendix:accuracy}

\begin{figure*}[ht]
    \centering
    \includegraphics[width=1.0\textwidth]{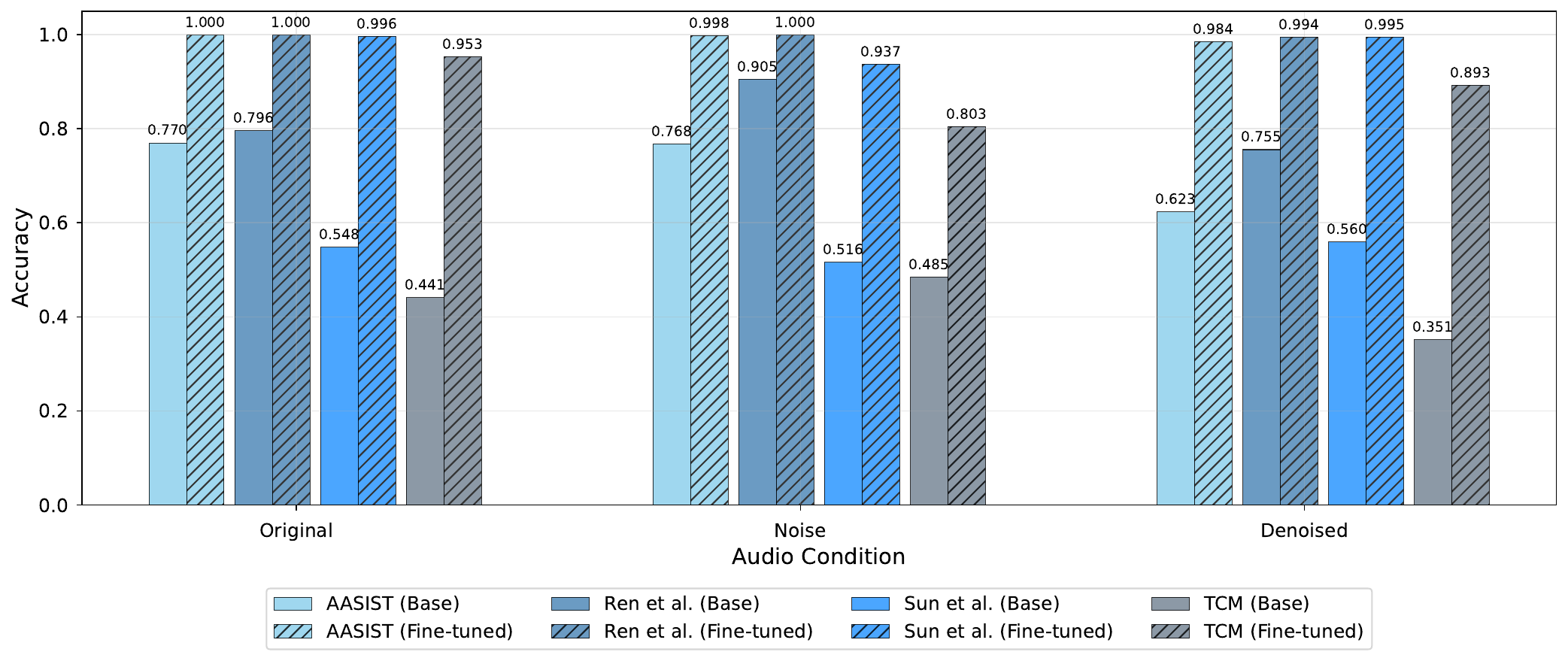}
    \caption{Accuracy of the four detectors on FOICE originals, noisy, and denoised audio. Solid bars show baseline performance, and hatched bars show fine-tuned variants. After fine-tuning, AASIST and Ren et al.~\cite{AISVGeneralization} achieve near-perfect accuracy across all conditions, while Sun et al.~\cite{VocoderArtifacts} and TCM improve substantially, though remain weaker overall. The results highlight the effectiveness of fine-tuning in boosting in-distribution detection, even under degraded audio conditions.}
    \label{fig:accuracy_combined_comparison}
\end{figure*}

Figure~\ref{fig:accuracy_combined_comparison} shows the classification accuracy of the four  detectors across FOICE originals, noisy, and denoised audio conditions. The results demonstrate  that FOICE fine-tuning consistently improves accuracy, with AASIST and Ren et al.~\cite{AISVGeneralization} achieving 
near-perfect performance, while Sun et al.~\cite{VocoderArtifacts} and TCM also benefit though remain less robust overall.

\section{Spectrograms of vocoders used to train baseline detectors.}\label{appendix:other_spectrograms}

\begin{figure*}[ht]
    \centering
    \includegraphics[width=1.0\textwidth]{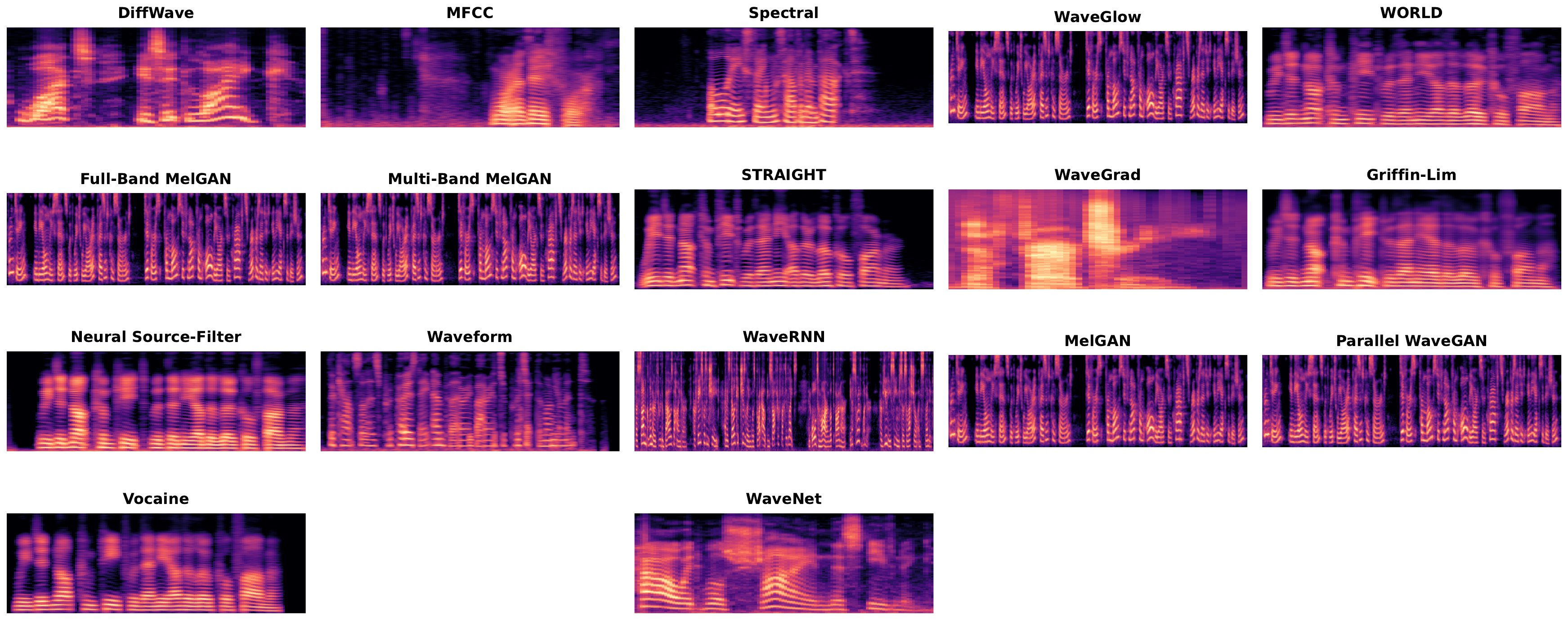}
    \caption{Spectrograms of vocoders present in LibriSeVoc, ASVspoof2019, and WaveFake, corresponding to the training sources of AASIST, Sun et al. \cite{VocoderArtifacts}, Ren et al. \cite{AISVGeneralization} (variant trained on LibriSeVoc and ASVspoof2019), and TCM; each vocoder displays unique spectrotemporal patterns that baseline detectors may overfit to or miss, depending on dataset composition.}
    \label{fig:combined_spectrograms}
\end{figure*}

The spectrograms in Figure~\ref{fig:combined_spectrograms} present representative samples from the vocoders contained in the datasets used to train the baseline detectors, with each vocoder exhibiting unique spectral envelopes, excitation characteristics, and artifact structures that are visually distinct in both frequency and time. The training provenance for each detector is as follows: AASIST and TCM were both trained using ASVspoof2019; Sun et al.~\cite{VocoderArtifacts} was trained on LibriSeVoc, ASVspoof2019, and WaveFake; and, following the protocol in \cite{AISVGeneralization}, Ren et al. \cite{AISVGeneralization} was trained in a leave-one-dataset-out manner on LibriSeVoc, ASVspoof2019, WaveFake, and FakeAVCeleb—we utilize the model version trained with LibriSeVoc and ASVspoof2019. The datasets themselves span a broad set of vocoders: LibriSeVoc incorporates DiffWave, MelGAN, Parallel WaveGAN, WaveGrad, WaveNet, WaveRNN; WaveFake includes MelGAN, Parallel WaveGAN, WaveGlow, MultiBand-MelGAN, and FullBand-MelGAN, HiFi-GAN; while ASVspoof2019 contains WaveNet, WaveRNN, Spectral, WORLD, Waveform, Griffin-Lim, MFCC-vocoder, Neural source-filter, STRAIGHT, and VocAine; for FakeAVCeleb, there is only SV2TTS.

\end{document}